# DIELECTRIC PROPERTIES OF VESTA'S SURFACE AS CONSTRAINED BY DAWN VIR OBSERVATIONS

Elizabeth M. Palmer[a,b], Essam Heggy[c,d], Maria T. Capria[e], Federico Tosi[e]

[a] Department of Earth, Planetary, and Space Sciences, University of California, Los Angeles, 595 Charles Young Drive East, Box 951567, Los Angeles, CA, USA 90095-1567.

[b] Department of Geosciences, Western Michigan University, 1903 W Michigan Avenue, MI, USA 49008-5241. (Elizabeth.M.Palmer@wmich.edu)

[c] Jet Propulsion Laboratory, California Institute of Technology, 4800 Oak Grove Drive, MS 300-243, Pasadena, CA, USA 91101-8099.

[d] University of Southern California, Ming Hsieh Department of Electrical Engineering, 3737 Watt Way, Los Angeles, CA 90089, USA (heggy@usc.edu)

[e] Institute for Space Astrophysics and Planetology, INAF-IAPS, via del Fosso del Cavaliere 100, 00133 Rome, Italy. (mariateresa.capria@iaps.inaf.it; federico.tosi@iaps.inaf.it)

**Corresponding Author:** [b] Elizabeth Palmer, Now at the Department of Geosciences, Western Michigan University, 1903 W Michigan Avenue, MI, USA 49008-5241, Elizabeth.M.Palmer@wmich.edu

Highlights:

- We establish the first surface dielectric model of asteroid Vesta

- The dielectric constant $\varepsilon'$ at Vesta's surface at X- and S-band radar frequencies is ~2.4

- The dielectric constant of Vesta is constant across its surface

- We estimate a near-surface porosity of ~55% from Dawn VIR data analysis

**Abstract**

Earth and orbital-based radar observations of asteroids provide a unique opportunity to characterize surface roughness and the dielectric properties of their surfaces, as well as potentially explore some of their shallow subsurface physical properties. If the dielectric and topographic properties of asteroid's surfaces are defined, one can constrain their surface textural characteristics as well as potential subsurface volatile enrichment using the observed radar backscatter. To achieve this objective, we establish the first dielectric model of asteroid Vesta for the case of a dry, volatile-poor regolith—employing an analogy to the dielectric properties of lunar soil, and adjusted for the surface densities and temperatures deduced from Dawn's Visible and InfraRed mapping spectrometer (VIR). Our model suggests that the real part of the dielectric constant at the surface of Vesta is relatively constant, ranging from 2.3 to 2.5 from the night- to day-side of Vesta, while the loss tangent shows slight variation as a function of diurnal temperature, ranging from $6 \times 10^{-3}$ to $8 \times 10^{-3}$. We estimate the surface porosity to be ~55% in the upper meter of the regolith, as derived from VIR observations. This is ~12% higher than previous estimation of porosity derived from previous Earth-based X- and S-band radar observation. We suggest that the radar backscattering properties of asteroid Vesta will be mainly driven by the changes in surface roughness rather than potential dielectric variations in the upper regolith in the X- and S-band.

## 1. Introduction

NASA's Dawn mission is targeting two uniquely large asteroids for orbital investigation, Vesta and Ceres, which are thought to be remnant building blocks of the terrestrial planets (Russell et al., 2011). The structural and textural properties of asteroids are observed primarily using Earth-based radar, and provide insight into the processes that shaped their surfaces: whether through impact cratering, lava flows, or stress fracturing as a result of the diurnal thermal erosion arising from the expansion and contraction of volatiles embedded in the surface material. The first target of the Dawn mission, Vesta, was expected to have depleted its volatile content long ago through global melting, differentiation, and later regolith gardening by impacts from smaller bodies (Russell et al., 2011). However, multiple observations from Dawn's year-long orbital mission point to the ephemeral presence of volatiles: localized hydrogen concentrations in regions thought to contain impactor-delivered hydrated materials (Prettyman et al., 2012; Reddy et al., 2012), widespread hydroxyl absorption bands across the surface (De Sanctis et al., 2012b), pitted terrain in some crater floors, thought to be caused by the degassing of subsurface volatiles (Denevi et al., 2012), and gullies in crater walls that are morphologically consistent with formation by transient fluid flow (Scully et al., 2015).

Given radar's ability to resolve a target's overall shape and centimeter- to decimeter-scale surface roughness, and its ability to assess the potential presence of ice through polarimetric ratios (e.g. Thompson, Ustinov & Heggy, 2011; Thompson et al., 2012), radar is a particularly useful technique for aiding in the volatile investigation of Vesta's surface and that of other small bodies. Earth-based and orbital radar studies of the Moon, for example, have revealed potential sites of ice concentration at the poles (e.g. Spudis et al., 2010), while Earth-based radar observations of Mercury have provided the first detection of water ice in permanently shadowed craters (Slade, Butler & Muhleman, 1992). For asteroids, Earth-based radar is predominantly used for detection purposes, yielding shape, spin and qualitative surface roughness from delay-Doppler imaging (Ostro et al., 2002). Vesta has likewise been observed at a number of radar frequencies: the X-band at Goldstone and S-band at Arecibo (e.g. Mitchell et al., 1996), as well as the Ku- and C-bands with the Very Large Array (VLA) (Johnston et al., 1989).

The results of such radar observations, however, are difficult to translate into quantifiable surface physical properties, as the power and polarization of the returned radar backscatter are affected by the observing geometry (which is easily constrained) and by three intrinsic factors of the surface: (1) variations in the surface topography, (2) variations in surface roughness, and (3) variations in the surface's dielectric properties—which describes the radar reflectivity and absorptivity of a material, and is primarily dependent on the mineralogy, bulk density, temperature and volatile content of the surface material (e.g. Heggy et al., 2001; Paillou et al., 2006; Heggy & Palmer et al., 2012). While the effect of the surface topography of small bodies can be modeled from a shape model (which in turn can be derived from speckle interferometry, photometric lightcurves or stereoscopic images), surface roughness and surface dielectric properties remain poorly characterized. As a consequence, when measuring the radar backscatter from the surface of a small body, it is challenging to determine whether backscatter variations are caused by variations in surface roughness or variation in the surface's dielectric properties. The ability to construct a quantifiable surface roughness map, and subsequently to identify regions that are smooth at decimeter scales, is critical to the success of future small-body landing missions (e.g. Asphaug 2006) and future sampling experiments (ElShafie & Heggy, 2013).

This difficulty is exemplified by Mitchell et al. (1996), who used radar Doppler spectra to qualitatively infer that Vesta's surface is overall rougher than the Moon at both decimeter and centimeter scales. While they used a shape model (derived from speckle interferometry) to constrain the topographic component of the total radar backscatter, they did not estimate the backscatter contribution that arises from potential variations in the surface's dielectric properties—which are widely used to assess the textural and compositional uncertainties of a surface (e.g. Boisson et al., 2009 & 2011). On the Moon, the ability to quantify dielectric properties has proven significant for identifying distinctions between the two types of lunar terrain, the highlands and the lunar *maria* (Fa & Wieczorek, 2012).

One study that attempts to estimate the dielectric properties of Vesta's surface is conducted by Johnston et al. (1989), who find that the Ku- and C-band microwave emissions from Vesta are in disagreement with those expected of a rotating blackbody. They suggest that the asteroid may be covered by a thin layer of dust that decreases microwave reflectivity, thereby increasing the body's microwave brightness. When estimating the thickness of this layer, they rely on dielectric mixing models of generic powdered rock (discussed by Campbell & Ulrichs (1969)), and suggest a depth of 6 cm when assuming a value of 2.9 for the real part of the relative dielectric constant $\varepsilon'$ and assuming $1.5\times10^{-2}$ for the loss tangent tan $\delta$—where $\varepsilon'$ relates to the material's reflectivity and tan $\delta$ to the material's attenuation of energy. Johnston et al.'s (1989) value of $\varepsilon'$, however, is inconsistent with dielectric laboratory measurements of powdered basaltic samples near the same bulk density of 1.00 g cm$^{-3}$ (e.g. Campbell & Ulrichs, 1969; Alvarez, 1974). When considered alongside the study of Mitchell et al. (1996), both results suggest that there is a substantial ambiguity regarding the textural and dielectric properties of Vesta's surface, as well as in the method used to quantify them from Earth-based radar observation.

With the Dawn mission, however, an opportunity arises to address this deficiency. In this study, thermal observations by Dawn's Visible and InfraRed (VIR) mapping spectrometer are used to constrain the bulk density and temperatures of the surface, which are the main parameters that determine the surface's dielectric properties for a desiccated planetary regolith (Thompson, Ustinov & Heggy, 2011). Our dielectric model is constructed specifically for the dry, volatile-poor case of Vesta's surface, given that the highest hydrogen concentration observed by Dawn's GRaND instrument is 400 ppm or 0.04 wt.% (Prettyman et al., 2012), which is well below the radar detectability limit of at least 10% of ice content in basaltic desiccated lunar-like soils (Fa, Wieczorek & Heggy, 2011). This model allows us to assess the expected range of dielectric properties arising from potential surface compositional variations (as described by De Sanctis et al., 2012a & 2013), as well as from variations in surface temperature and density (as described by Tosi et al., 2014 & Capria et al., 2014).

Since the dielectric properties of asteroid analog materials have yet to be measured in the laboratory, we use existing dielectric studies of lunar soil samples to serve as suitable analogs to Vesta's upper regolith material. The compositional analogy between Vesta's surface material and basaltic lunar soil is addressed in Sections 2.1 and 2.2, and the assumptions and limitations of the resulting surface dielectric model are considered in Sections 2.3 and 2.4. Section 3 contains the results of the dielectric model for Vesta's surface, and includes suggested validation sites where the dielectric constant of Vesta's surface may be estimated from future in-situ radar observations. In Section 4, the implications of our findings are discussed for existing and future Earth- and space-based radar studies of Vesta and other asteroids, with emphasis on volatile detectability. Overall, this dielectric model is an essential step toward retrieving quantifiable surface roughness from radar backscatter measurements by Earth-based and orbital X- and S-band radar observations (e.g. Mitchell et al., 1996; Nolan et al., 2005), which will develop understanding of the processes that govern Vesta's surface texture, as well as support the planning of potential future landing and sampling experiments, such as NASA's Origins-Spectral Interpretation-Resource Identification-Security-Regolith Explorer (OSIRIS-REx) (Lauretta et al., 2014).

## 2. Dielectric model construction and parameter constraint

The dielectric properties of a material describes the intrinsic mechanisms by which the material reflects and attenuates the electric field component of the incident radar wave, and is quantified by the complex relative dielectric constant ($\varepsilon = \varepsilon' + i\varepsilon''$). For brevity, the relative dielectric constant, a dimensionless quantity, is hereon referred to as the dielectric constant. As previously mentioned, the real part of the dielectric constant, $\varepsilon'$, relates to the reflectivity of the material, while the ratio of the imaginary part to the real part ($\varepsilon''/\varepsilon'$) is termed the loss tangent (tan $\delta$), and quantifies the loss of energy during transmission through a given material (such as Vesta's regolith). For small desiccated terrestrial bodies, including the Moon and Mercury, the dielectric properties of the surface mainly depend on the material's (1) mineralogy, (2) bulk density, (3) diurnal surface temperature, (4) potential volatile content (e.g. Heggy & Palmer et al., 2012), and (5) frequency, which falls in the range of 2–8 GHz for Earth-based radar observations. In the following sections, the first four of these geophysical parameters are constrained for the material of Vesta's upper regolith using observations by the Dawn VIR spectrometer. In Section 3, these constraints are utilized to construct a numerical model of $\varepsilon'$ and tan $\delta$ for the general case of a volatile-poor, dry surface and shallow subsurface of Vesta.

### 2.1. Surface mineralogy

Vesta has been identified as the parent body of basaltic, achondritic meteorites (howardites, eucrites and diogenites; "HEDs") through extensive meteoritic studies and Earth-based spectral observations of asteroids (Takeda, 1997; Hiroi, Pieters & Takeda, 1994). As confirmed by hyperspectral observations from Dawn's VIR instrument (De Sanctis et al., 2012a & 2013), the composition of Vesta's upper regolith is analogous to that of howardite—a brecciated material formed by clasts of eucrite and diogenite in varying ratios.

The surface contains a heterogeneous distribution of eucrite-rich versus diogenite-rich howardite (e.g. De Sanctis et al., 2013), where the primary mineralogical difference between eucrites and diogenites is in the pyroxene composition and content. Eucrites are primarily composed of low-Ca pyroxene and plagioclase, while diogenites contain a high abundance of low-Ca, Mg-rich pyroxene with only minor amounts of plagioclase (e.g. Burbine et al., 2001).

The above-described surface mineralogy is expected to provide minimal dielectric variation on the surface of Vesta. For instance, dielectric measurements of lunar samples by Olhoeft and Strangway (1975)—consisting of the same major minerals as HEDs at varying concentrations—suggest that the real part of the dielectric constant ($\varepsilon'$) is mainly a function of the bulk density, and shows no substantial variation with composition among the lunar basaltic minerals commonly found in HEDs. While the loss tangent of lunar samples varies with ilmenite content (Carrier, Olhoeft & Mendell, 1991), howardite, eucrite and diogenite samples in the NASA Johnson Space Center's HED meteorite compendium each contain only minor, accessory amounts of ilmenite ($\leq 1.5\%$) (Righter & Garber, 2011). The above suggests that Vesta's surface can be expected to be dielectrically homogeneous when considering the effect of composition on the dielectric properties of the regolith. Furthermore, given the extensive gardening of regolith by meteoritic impacts, the upper few meters of the Vestan regolith are also expected to be compositionally homogeneous with depth (Pieters et al., 2012) and hence the effect of composition on vertical dielectric variation in the first few meters can be neglected.

## 2.2. Compositional analogy to lunar soil

While Vesta's surface mineralogy is consistent with howarditic dust, howardite has yet to be dielectrically characterized through laboratory study. In order to estimate Vesta's dielectric properties for given conditions of surface density and temperature, one must identify an analog material that has been sufficiently dielectrically characterized under Vesta's relevant surface conditions.

We hypothesize that basaltic lunar soil is the most suitable compositional analog to the Vestan regolith. Both lunar soil and regolithic howardites are brecciated basalts—gardened by meteoritic impacts and composed primarily of pyroxenes and plagioclase (Papike, Taylor & Simon, 1991; Warren et al., 2009)—and originate from airless, desiccated regoliths unlike Earth or martian basalts. Furthermore, lunar soil samples have been measured extensively in the laboratory for dielectric variation over a wide range of radar frequencies, for various bulk densities (e.g. Olhoeft & Strangway, 1975), at low temperatures, and in vacuum, relevant to planetary conditions (e.g. Alvarez, 1974).

Lunar soil and regolithic howardite differ in their minor mineralogy (Cartwright et al., 2013), but unless one such mineral has a significantly high dielectric constant (e.g. iron oxide minerals) compared to that of the major host minerals, its presence does not measurably alter the bulk material's dielectric constant (Campbell et al., 2002). Ilmenite content in lunar soils, for example, will increase the loss tangent of the material with increasing volume fraction (Carrier, Olhoeft & Mendell, 1991). Among the collection of HED meteorites curated by NASA Johnson

Space Center (JSC), however, total ilmenite content is less than 1.5% and is typically found as only an accessory mineral in HED meteorites (Righter & Garber, 2011). The lunar samples selected for this study (Section 2.3) are also low in ilmenite content (<1.5%) (Meyer, 2010a, 2010b; Hill et al., 2007), further supporting the analogy between the dielectric properties of lunar soil with that of Vesta's surface material.

Vesta's regolith also undergoes different space weathering than on the Moon, such that Vesta's regolith has a lower noble gas content (Cartwright et al., 2013), and lacks nanophase iron (Pieters et al., 2012). Noble gas content, however, provides a negligible contribution to the dielectric properties of a soil (Olhoeft & Strangway, 1975), and Barmatz et al. (2012) find no dependence of dielectric properties on of the fraction of nanophase iron in lunar soil samples.

For the construction of the dielectric model of Vesta's surface, we hence conclude that basaltic lunar soil is a suitable analog to the bulk composition of Vesta's upper regolith.

*2.3. Surface density and diurnal temperature variation*

As discussed above for Vesta's surface, the spatial and temporal variation of the dielectric constant is expected to be primarily dependent on bulk density and secondarily dependent on temperature. In order to constrain the surface density and diurnal temperature variation of Vesta's regolith, we use the results of Capria et al. (2014), who model heat transfer in the upper 50 meters of Vesta's regolith by balancing solar input energy with output surface radiation. The Capria et al. (2014) thermophysical model accounts for (1) the thermal conductivity profile of the upper regolith (which depends on the best-fit density profile and temperature profile in the subsurface), as well as (2) the local sub-pixel surface topography (which suppresses the output surface radiation with increasingly rough terrain, and leads to an increase in local surface temperature).

Capria et al. (2014) ultimately deduce surface densities that provide the best fits between theoretical diurnal surface temperatures, and ones directly retrieved from VIR observations (Tosi et al., 2014). Their results are expressed as a map of regionally-averaged thermal inertia values, which quantify the resistance of a material to variations in its diurnal temperature (Capria et al., 2014). Regions with the most diurnal temperature change correspond to surfaces of low thermal inertia, which can be explained by fine, low-density material. While VIR is only sensitive to temperatures $\geq$180 K (Tosi et al., 2014), the thermophysical model of Capria et al. (2014) estimates the full range of day-to-night surface temperatures throughout a diurnal cycle. Temperatures retrieved from VIR observations ($\geq$ 180 K) are used to constrain daytime temperatures, while ESA's Herschel IR observations provide constraints on cooler and nighttime temperatures (Leyrat et al., 2012). Capria et al.'s (2014) thermophysical model additionally provides a best-fit surface density for each modeled region of Vesta's surface.

On average, Capria et al. (2014) find a best-fit surface density of ~1.30 g cm$^{-3}$, with local variations on the order of ± 0.10 g cm$^{-3}$. For the purpose of this study, in which surface bulk density and surface temperatures are used as input to the surface dielectric model, Capria (this study) provides a global map of theoretical surface temperatures for a given time in Vesta's rotation (Figure 1) using the globally-averaged surface density of 1.30 g cm$^{-3}$. A full description of the procedure for determining best-fit surface densities and diurnal temperature curves is detailed by Capria et al. (2014).

*2.4. Dielectric properties of Vestan analog material*

In Section 2.2, we proposed basaltic lunar soil to be a suitable compositional analog to Vesta's surface material, and in Section 2.3 used VIR observations to determine a range of surface bulk densities and temperatures over which to provide an estimate of the dielectric properties of Vesta's upper regolith. Three dielectric studies that sufficiently meet our criteria for relevant measurement conditions are those of Alvarez (1974), Frisillo, Olhoeft & Strangway (1975), and Bussey (1979), who each conducted dielectric laboratory studies on basaltic lunar fines from Apollo 17 (samples 74241, 72441 and 70051, respectively) under vacuum to ensure a totally desiccated sample, at temperatures below 300 K, and with sample bulk densities between ~1.30 g cm$^{-3}$ and ~1.80 g cm$^{-3}$. Unfortunately, few other Apollo lunar soil samples were protected against atmospheric moisture contamination, which measurably alters the dielectric constant (Olhoeft, Strangway & Pearce, 1975; Heggy et al., 2001), and few were measured in vacuum conditions, limiting the use of additional lunar sample dielectric studies for the Vestan case.

While the dielectric measurements of Alvarez (1974), Frisillo, Olhoeft & Strangway (1975) and Bussey (1979) have each been conducted at frequencies below the 2–8 GHz range as used by Earth-based radar observations, desiccated basalts that lack iron-oxide minerals are non-dispersive dielectric materials over the MF, HF, VHF, UHF and SHF radio frequency range, with no observed relaxations over 0.1 MHz to 8 GHz when performing a dielectric spectroscopic characterization (Carrier, Olhoeft & Mendell, 1991; Heggy et al., 2007; Brouet et al., 2014). Since (1) the modal content of ilmenite, an iron oxide mineral common in lunar regolith, is less than 1.5% in each of the three lunar soil samples, and also in each of the HED meteorites in the NASA JSC collection, and (2) samples are totally desiccated, there is no sufficient presence of conductive ions to cause a measurable frequency dependence behavior in both the real and imaginary part of the dielectric constant as measured across the dipolar polarization regime of the electromagnetic spectrum. One can therefore safely apply the dielectric measurements of Alvarez (1974) at 0.1 MHz, of Frisillo, Olhoeft & Strangway (1975) at 0.1 MHz, and of Bussey (1979) at 1.6 GHz to X- and S-band radar frequencies.

Table 1 summarizes the dielectric measurements of the three selected lunar soil samples. Alvarez (1974) compacted soil sample 74241 to a bulk density of 1.38 g cm$^{-3}$, and under vacuum conditions, measured $\varepsilon'$ and tan $\delta$ at successive temperatures of 100, 298 and 373 K. The procedure was repeated when the sample was compacted to 1.61 g cm$^{-3}$. At each given temperature, $\varepsilon'$ and tan $\delta$ were measured in the frequency range of 30 Hz to 0.1 MHz. Values reported in Table 1, and used in the construction of our dielectric model, correspond to those measured at the maximum frequency of 0.1 MHz, at which $\varepsilon'$ ranges from 2.18 to 2.42, and tan $\delta$ ranges from $1.2 \times 10^{-2}$ to $1.7 \times 10^{-2}$.

The dielectric measurements of Frisillo, Olhoeft & Strangway (1975) were conducted similarly on lunar soil sample 72441 at sample bulk densities of 1.56, 1.65 and 1.80 g cm$^{-3}$. Each measurement of $\varepsilon'$ and tan $\delta$ were conducted under vacuum, over the frequency range of 100 Hz to 0.1 MHz, and at temperatures of 298, 323, and 373 K. Their measurements yielded $\varepsilon'$ between 3.04 and 3.27, and tan $\delta$ between $4 \times 10^{-3}$ and $7 \times 10^{-3}$. Again, values of $\varepsilon'$ and tan $\delta$ reported in Table 1 correspond to the maximum frequency of measurement by Frisillo, Olhoeft & Strangway (1975) at 0.1 MHz.

In the third study, Bussey (1979) measured the dielectric properties of sample 70051 at a fixed bulk density of ~1.60 g cm$^{-3}$ over the frequency range of 500 MHz to 10 GHz, and at temperatures of 173, 232, 296 and 373 K under vacuum. The dielectric properties reported in Table 1, however, correspond to measurements at 1.6 GHz rather than 10 GHz, as Bussey (1979) states that measurements above this threshold were unreliable due to instrumental error. Bussey's

(1979) measurements of $\varepsilon'$ and tan $\delta$ for sample 70051 range from 3.30 to 3.42 and $5\times10^{-3}$ to

$8\times10^{-3}$, respectively.

In order to extrapolate the measurements of Alvarez (1974), Frisillo, Olhoeft &

Strangway (1975) and Bussey (1979) to the relevant densities and temperatures of Vesta's

regolith, we apply an iterative, multivariable least-squares fit to the data in Table 1 to determine empirical formulas for both the real part $\varepsilon'$ of the dielectric constant and the loss tangent tan $\delta$ of the lunar samples as functions of bulk density and temperature (Markwardt, 2009). We fit a power-law to the density dependence of $\varepsilon'$, and find a minor dependence of $\varepsilon'$ on temperature,

consistent with dielectric laboratory measurements of other lunar samples (e.g. Olhoeft &

Strangway, 1975). The loss tangent of the lunar samples follows an exponential dependence on

temperature, and exhibits no apparent density dependence between 1.38 g cm$^{-3}$ and 1.80 g cm$^{-3}$

among the three selected lunar samples.

The resulting best-fit empirical formulas to the dielectric properties of the three selected

lunar soil samples are as follows:

$$\varepsilon'(\rho,T) = \left(1.85 \pm 0.02\right)^{\rho} + \left(8 \pm 1\right)\times10^{-4}T \tag{1}$$

$$\tan\delta(T) = (5.1 \pm 0.2)\times10^{-3}\exp\{(1.7 \pm 0.2)\times10^{-3}T\} \tag{2}$$

where $\varepsilon'$ is the real part of the dielectric constant of dry, basaltic soil that lacks significant iron-oxide content; tan $\delta$ is its loss tangent; $T$ (K) is the soil's temperature; and $\rho$ (g cm$^{-3}$) is the bulk

density of the soil. This dielectric model is applicable over the range of ~1.20 g cm$^{-3}$ to 1.80 g

cm$^{-3}$ for desiccated basaltic soil, as constrained by the range of bulk densities measured in the dielectric studies of Alvarez (1974), Frisillo, Olhoeft & Strangway (1975) and Bussey (1979).

The base of the exponent (1.85) in Equation 1 is comparable to that reported by Olhoeft &

Strangway (1975), who compiled 92 measurements of the dielectric constant vs. density of lunar samples, and found the base of the exponent to be $1.93 \pm 0.17$. Notably, their selection criteria were less strict, as they included many studies of lunar soils that were measured under ambient atmosphere, and therefore exposed to moisture contamination, which measurably impacts both $\varepsilon'$ and $\tan \delta$ (Olhoeft, Strangway & Pearce, 1975; Heggy et al., 2001). The small temperature dependence and lack of density dependence in the loss tangent (Equation 2) also agree with the findings of Olhoeft & Strangway (1975). Using these empirical formulas as estimates for the dielectric properties of Vesta's surface, maps of $\varepsilon'$ and $\tan \delta$ are constructed for a given time in the asteroid's rotation (displayed in Figure 2) and use, as inputs, the modeled surface temperatures of Vesta from Figure 1, which range from 98 to 280 K, and the average best-fit surface bulk density of 1.30 g cm$^{-3}$.

### 3. Dielectric model of a volatile-free surface

*3.1. Average near-surface dielectric properties*

Figure 2 shows the resulting global model of $\varepsilon'$ (top panel) and $\tan \delta$ (bottom panel) for Vesta's surface at a spatial resolution of 5° latitude by 10° longitude per pixel. Along the equator, this corresponds to a spatial resolution of ~5 by 10 km/pixel. Illumination conditions during the *Approach Phase* of the Dawn mission limit the derivation of surface albedos to latitudes between 70°S and 30°N (Li et al., 2013), and thereby confine both the thermal inertia calculations of Capria et al. (2014) and the dielectric model in this study to the same latitude range.

For X- and S-band radar observations, we report two sets of dielectric properties: (1) those of the surface and near-surface (i.e. ~1-2 wavelengths deep), which corresponds to the surface average bulk density of ~1.30 g cm$^{-3}$, and (2) the dielectric properties in the potential case of penetration depth of up to ~10 wavelengths, which corresponds to the upper meter of regolith (Campbell, 2002; Thompson, Ustinov & Heggy, 2011). We use the same subsurface density-depth profile that was hypothesized in the thermophysical model of Capria et al. (2014), which is that of the lunar regolith given by Carrier, Olhoeft & Mendell (1991). This yields a maximum bulk density of ~1.80 g cm$^{-3}$ at one meter depth.

Using these two bulk densities as inputs to Equation 1, and accounting for the VIR observed range of diurnal temperatures at the surface ranging from 98 to 280 K, $\varepsilon'$ is calculated to vary from 2.31 to 2.47 at the surface, and to vary from ~3.10 to 3.25 at one meter depth. For the loss tangent, the diurnal effect leads to a range of $6\times10^{-3}$ to $8\times10^{-3}$ respectively between the night and day side.

In regions of anomalously high or low thermal inertia values, which Capria et al. (2014) attribute mainly to a difference in surface bulk density ($\pm$ 0.10 g cm$^{-3}$), $\varepsilon'$ is expected to deviate $\leq$ 9% from the mean, based on the empirical model in Equation 1. At the relatively coarse spatial resolution of both the temperature and associated dielectric maps, these deviations are smoothed out, and the average surface density of 1.30 g cm$^{-3}$ can be considered uniform in the construction of our surface dielectric model.

*3.2. Vesta's surface dielectric calibration sites*

In order to provide means for future validation of this dielectric model, four sites of ~1 km$^2$ are identified on the Vestan surface that have low roughness at ~20 m resolution, as defined from high-resolution imagery acquired by Dawn's framing camera (FC) during Low-Altitude Mapping Orbit (LAMO). Figure 3 shows the high-resolution framing camera images of the validation sites and their locations with respect to Vesta's global map. Figure 4 shows surface temperatures in each location as retrieved from VIR thermal imagery, as well as their surface dielectric properties as assessed from our empirical model in Equations 1 and 2.

Each site is consistent with a location of accumulation of fine-grained material, and is, furthermore, likely to be texturally smooth to the scale of meters (Jaumann et al., 2012). Radar backscatter from planetary regolith depends on both the surface roughness and the material's dielectric constant (e.g. Fa, Wieczorek & Heggy, 2011; Hagfors, 1964), hence by selecting sites of low surface roughness—characteristically flat, crater-free and smooth (i.e. lacking shadows)—this minimizes the ambiguity associated with deriving the surface's dielectric properties from the observed radar backscatter, as they must be disentangled from the scattering effects of surface roughness using inversion models (Tabatabaeenejad & Moghaddam, 2009) . While Earth-based radar observations of Vesta are constrained to an imaging resolution on the order of 10-100 km (Nolan et al., 2005), and are thereby unable to resolve such sites, future flyby and orbital missions to Vesta, for instance, can estimate the surface's dielectric constant from bistatic radar observations of the backscatter from these low-roughness locations (e.g. Simpson, 2011).

Site 4 is located in a region consistent with ponded crater ejecta material (labeled "Zone B" in Figure 3), and is centered on (22°S, 268°E) based on longitudes expressed in the Dawn Claudia coordinate system (Russell et al., 2012; Roatsch et al., 2012; Li et al., 2013). Surface temperatures in Zone B were retrieved from two VIR infrared images (~170 m/pixel resolution) that overlapped Site 4, and range from ~210-265 K. Using the average best-fit surface density of Capria et al. (2014) and the lunar density-depth profile of Carrier, Olhoeft & Mendell (1991), our dielectric model suggests that at the surface, $\varepsilon'$ is relatively constant at 2.4, and tan $\delta \sim 7.6 \times 10^{-3}$ to $8.0 \times 10^{-3}$. For the upper meter of regolith, $\varepsilon'$ is relatively constant at ~3.2 with the same loss tangent as at the surface due to its density invariance.

Calibration Sites 1-3 are located on the southwestern terrace in the wall of Marcia crater, designated "Zone A" in Figure 3. Surface temperatures within Zone A were retrieved from two VIR infrared images (~170 m/pixel resolution) that overlapped the sites, and range from ~220-255 K. De Sanctis et al. (2015) describe the observed temperatures of this region in detail as inferred from VIR, and Capria et al. (2014) find that the region's thermal inertia is slightly higher when compared to the global average for Vesta—potentially indicative of higher local surface density (on the order of ~1.32 g cm$^{-3}$) or coarser regolith. Figure 4 shows Sites 1-3 in the case of the average surface density of 1.30 g cm$^{-3}$, for which $\varepsilon' \sim$ 2.43 to 2.46 at the surface and $\varepsilon'$ is relatively constant at ~3.2 in the upper meter of Vesta's regolith; tan $\delta \sim 7.5 \times 10^{-3}$ to $7.9 \times 10^{-3}$ at both depths. If the local surface density is ~0.02 g cm$^{-3}$ higher than the average, $\varepsilon'$ at the surface is expected to be only ~1% greater than the global average at the surface.

## 4. Implications for characterizing Vesta's surface from radar observations

The first implication of this study pertains to constraining the ambiguities of Vesta's surface dielectric properties as assessed from Earth- and space-based radar backscatter measurements. For example, the received radar circular polarization ratio (CPR) describes the ratio between left- and right-hand polarizations of a surface's backscattered signal, and is often used to infer surface roughness and volatile enrichment (e.g. Thompson, Ustinov & Heggy, 2011; Thompson et al., 2012). Variations in the surface dielectric constant can also impact the CPR measurements and hence compromise the detectability of potential volatile presence (Fa, Wieczorek & Heggy, 2011). The dielectric model presented in this study suggests that the dielectric constant of Vesta's surface varies minimally with temperature, hence we conclude that observed CPRs will be primarily driven by surface roughness at the scale of the radar wavelength (Thompson et al., 2011). For regions with higher or lower values of thermal inertia relative to the average (such as within Marcia crater, and indicative of a difference in either regolith coarseness or surface density), $\varepsilon'$ is expected to differ by no more than ~9% from regional values, therefore not inducing a measurable change in CPR based on the analytical polarimetric radar scattering model established by Fa, Wieczorek & Heggy (2011). We therefore expect that observed variations in the CPR of Vesta's surface will correspond to changes in surface roughness rather than in dielectric properties.

The second implication of this study pertains to estimating the porosity of the upper meter of the Vestan regolith, which can be constrained by measuring the dielectric constant from radar observations. This is of particular interest for future landing and sample return missions to asteroids, as exemplified by NASA's OSIRIS-REx sample-return mission to asteroid Bennu (Lauretta et al., 2014), since the accurate determination of near-surface porosity is essential to identifying locations of low hardness in the regolith, whether to maximize the stability of platform anchoring or to optimize drilling rates for subsurface excavation (ElShafie & Heggy, 2013). Previous estimation of the dielectric constant of the Vestan regolith was conducted by Johnston et al. (1989), whose observations suggested a value of $\varepsilon' = 2.9$ and tan $\delta = 1.5 \times 10^{-2}$ for the upper 6 cm of the regolith. Our analysis, on the other hand, suggests $\varepsilon' \sim 2.3$ to $2.5$ at the surface and tan $\delta \sim 6.0 \times 10^{-3}$ to $8.3 \times 10^{-3}$. While Johnston et al. (1989) similarly adopt a lunar analogy for the textural and physical properties of Vesta's upper regolith, their dielectric constant would require a near-surface bulk density of ~1.60 g cm$^{-3}$ to achieve $\varepsilon' = 2.9$, based on the density-dependence of the dielectric constant of the lunar sample used in this study at $T = 280$ K (Equation 1). Assuming a solid basalt crustal density of ~2.90 g cm$^{-3}$ for Vesta (Raymond et al., 2011), and using our dielectric model to derive bulk density from Johnston et al.'s (1989) estimation of $\varepsilon'$, their results suggest a near-surface porosity of ~43%, whereas the best-fit surface bulk density of 1.30 g cm$^{-3}$ (Capria et al., 2014) suggests a porosity of ~55%. The latter is consistent with estimation by Magri et al. (2001), who use hyperspectral and radar observations of asteroid Eros as ground-truth to then estimate near-surface porosities on other small bodies; they find a near-surface porosity of $\sim 59\%^{+19}_{-11}$ for Vesta.

## 5. Conclusions and future work

We have established a surface dielectric model for the case of a volatile-free upper regolith on Vesta. This model depends on the bulk density and temperature of the upper regolith material as constrained by VIR observations, and the hypothesis that lunar basaltic soil is a suitable compositional analog to Vesta's regolith, as supported by Cartwright et al. (2013). At the surface, our model suggests that $\varepsilon'$ ranges from ~2.3 to 2.5 and tan $\delta$ from $6\times10^{-3}$ to $8\times10^{-3}$ in the X- and S- radar frequency bands; at one-meter depth, using a lunar density-depth profile, $\varepsilon'$ is expected to be ~3.2. For those regions of the Vestan surface exhibiting higher or lower thermal inertia than the global average (Capria et al., 2014), $\varepsilon'$ varies less than ±9% from the global average of ~2.4 at the surface, and ±3% from ~3.2 at one meter depth. This implies a near-surface bulk porosity ranging from 55% to 59% and decreasing to ~37% at one meter depth.

It is crucial to obtain accurate assessment of the surface's dielectric properties for asteroids and icy moons from future orbital radar missions to constrain the near-surface mechanical properties (e.g. porosity, density and compressive strength), and subsequently minimize risks associated with future landing, trafficability and coring activities for sample collection and analysis. For instance, passive bistatic radar observations are proposed to be part of Ganymede science observations in the JUICE mission, with the intent to characterize the surface's roughness, dielectric properties and surface porosities (Grasset et al., 2013). We suggest that the radar backscattering properties of asteroid Vesta will be mainly driven by the changes in surface roughness rather than potential dielectric variations in the upper regolith in the X- and S-band.

Future work will consist of combining opportunistic radar observations by Dawn's communications antenna with our dielectric model of Vesta's surface in order to produce surface roughness maps, which in turn can be used to help identify the various cratering processes that have shaped the surface.

## 6. Acknowledgements

The authors acknowledge Prof. Christopher Russell from UCLA for useful discussions, and also the Dawn Science, Instrument and Operations Teams for their support. Part of this research was carried out at the Jet Propulsion Laboratory, California Institute of Technology, under a contract with the National Aeronautics and Space Administration and was supported in part by the NASA Planetary Geology and Geophysics Program under grant NNXZ08AKA2G. VIR is funded by the Italian Space Agency (ASI) and was developed under the leadership of INAF-Istituto di Astrofisica e Planetologia Spaziali, Rome-Italy. Palmer was funded by a graduate student research award from UCLA.

**Table 1.** Dielectric properties of lunar soil samples used as compositional analogs to model the dielectric properties of Vesta's surface.

| Sample | Reference | Frequency | Bulk Density (g cm$^{-3}$) | Temperature (K) | $\varepsilon'$ | tan $\delta$ | Ilmenite Content (wt.%) |
|--------|-----------|-----------|------------------|-----------------|------|--------|------------------|
| 74241 | Alvarez (1974) | 100 kHz | 1.38 | 100 | 2.18 | 0.012 | 1.3[a] |
| | | | " | 298 | 2.20 | 0.014 | " |
| | | | " | 373 | 2.28 | 0.016 | " |
| | | | 1.61 | 100 | 2.34 | 0.012 | " |
| | | | " | 298 | 2.38 | 0.015 | " |
| | | | " | 373 | 2.42 | 0.017 | " |
| 72441 | Frisillo, Olhoeft & Strangway (1975) | 100 kHz | 1.56 | 298 | 3.04 | 0.004 | 0.3[b] |
| | | | " | 323 | 3.05 | 0.004 | " |
| | | | " | 373 | 3.09 | 0.005 | |
| | | | 1.65 | 298 | 3.13 | 0.004 | " |
| | | | " | 323 | 3.13 | 0.005 | " |
| | | | " | 373 | 3.16 | 0.006 | |
| | | | 1.8 | 298 | 3.27 | 0.005 | " |
| | | | " | 323 | 3.27 | 0.005 | " |
| | | | " | 373 | 3.28 | 0.005 | |
| 70051 | Bussey (1979) | 1.6 GHz | 1.6 | 173 | 3.30 | 0.005 | 2.8[c] |
| | | | " | 232 | 3.34 | 0.006 | " |
| | | | " | 296 | 3.42 | 0.007 | " |
| | | | " | 373 | 3.42 | 0.008 | " |

[a] Meyer (2010a)

[b] Meyer (2010b)

[c] Hill et al. (2007)

**Figure 1.** Global model of surface temperatures on Vesta. The map has a resolution of 5° latitude by 10° longitude per pixel. The map corresponds to an average best-fit surface density of 1.30 g cm$^{-3}$ with a sub-solar point of (26.7°S, 160.6°E) at a Vesta-Sun distance of ~2.1 AU. Modeled temperatures range from 98 to 280 K, where those below 180 K bear the most uncertainty, on the order of tens of K. Results are given from 70°S to 30°N due to the limited availability of albedo values derived from *Approach Phase* data.

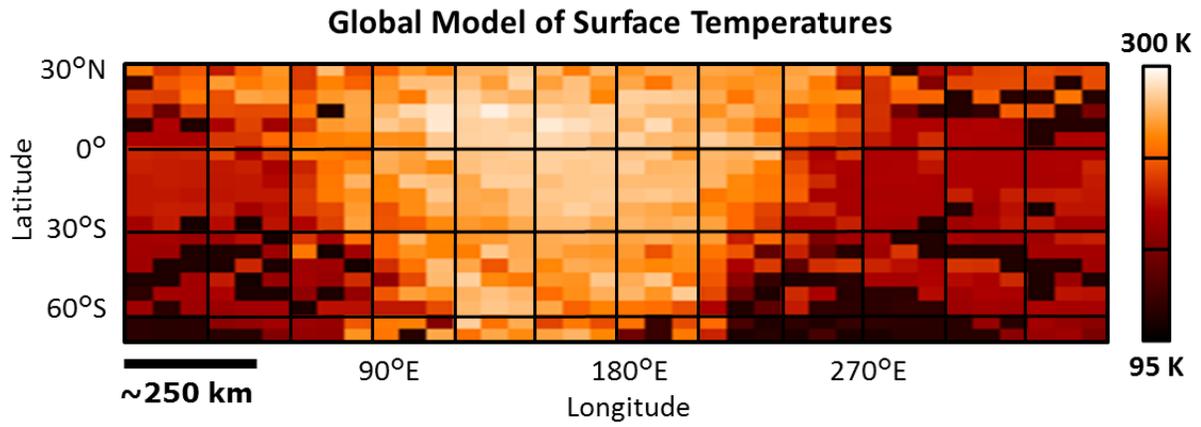

**Figure 2.** Global dielectric model of $\varepsilon'$ (top) and tan $\delta$ (bottom) on Vesta's surface. Results are applicable at X- and S-band frequencies, where each map has a resolution of 5° latitude by 10° longitude per pixel. Calculations assume a best-fit surface density of 1.30 g cm⁻³. The sub-solar point is at (26.7°S, 160.6°E) with a Vesta-Sun distance of ~2.1 AU. Results are evaluated between 70°S and 30°N due to the availability of *Approach Phase* data, where $\varepsilon'$ ranges between 2.3 and 2.5, and tan $\delta$ from $6.0\times10^{-3}$ to $8.3\times10^{-3}$. In areas of lower thermal inertia (potentially lower density) below the resolution of this map, our model suggests $\varepsilon' \sim 2.2$ to 2.3, and in areas of higher thermal inertia (higher density), $\varepsilon' \sim 2.3$ to 2.5.

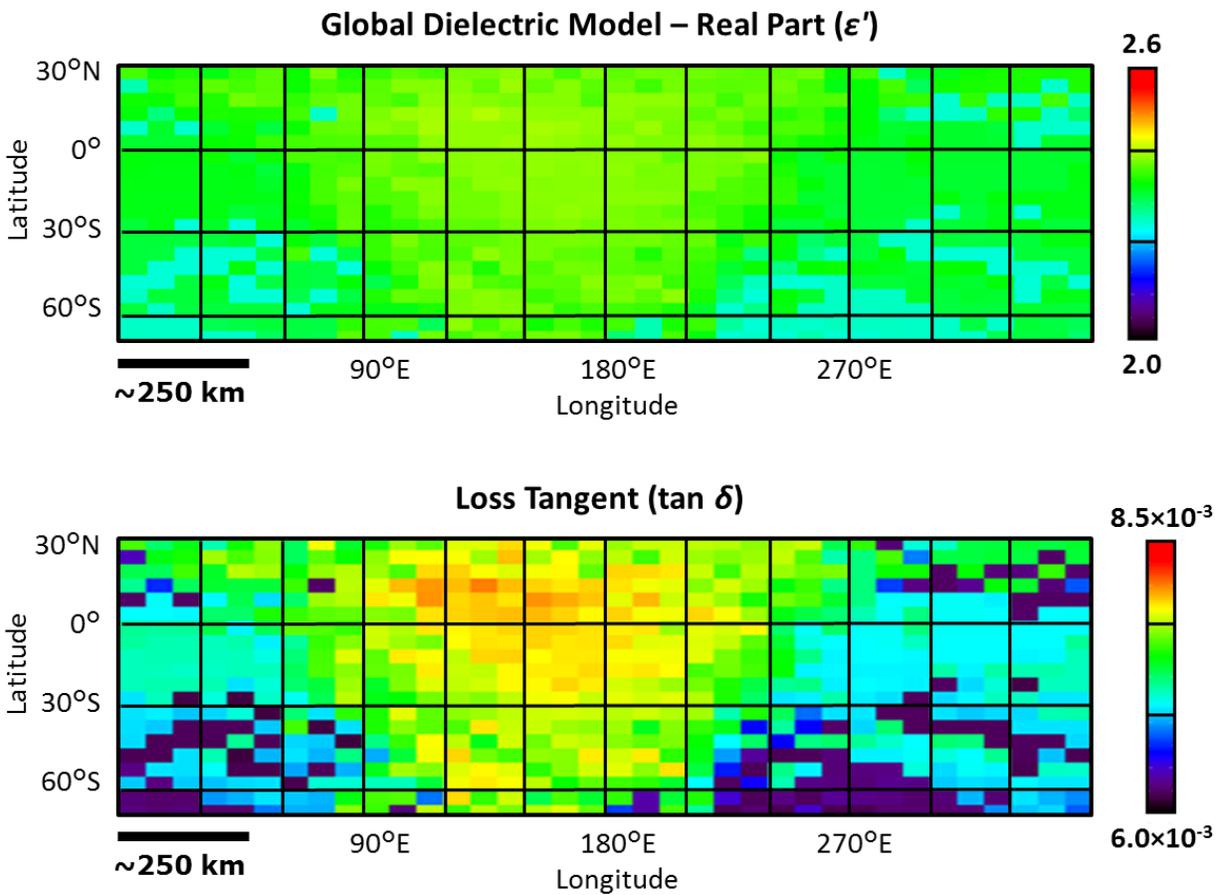

**Figure 3.** Potential dielectric calibration sites on Vesta's surface. The top-most image is a mosaic map of high-resolution FC images obtained at ~20 m/pixel—centered on (10°S, 230°E) in the Claudia crater coordinate system—and provides broader context for the location of the suggested calibration sites. Zone A (middle image) contains Sites 1-3 and is centered on (5°N, 187°E). Zone B (bottom image) contains Site 4 and is centered on (22°S, 265°E). Each site is characterized by a lack of shadows, and is consistent with a location of accumulation of fine-grained material. Future flyby or orbital radar observations of these low-roughness sites may be used to estimate the surface's dielectric constant.

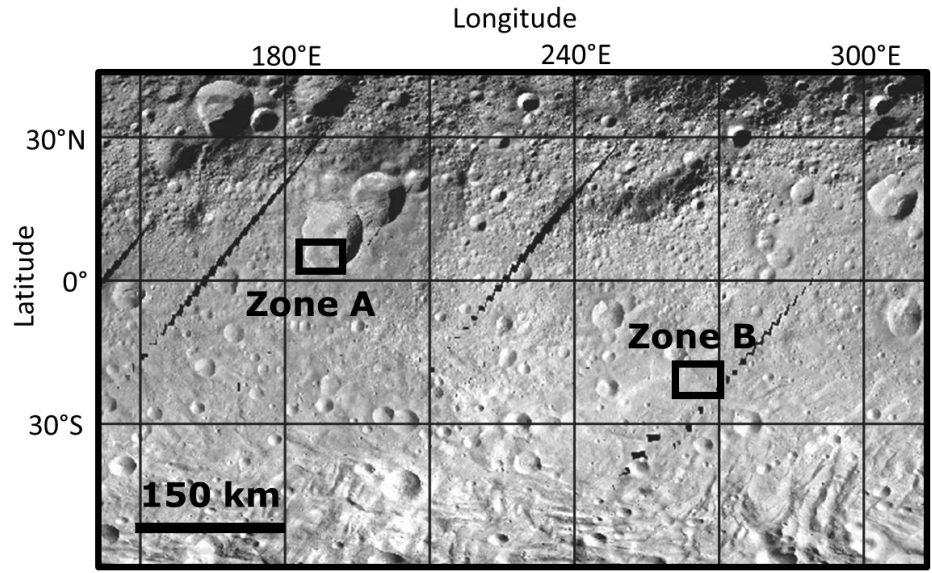

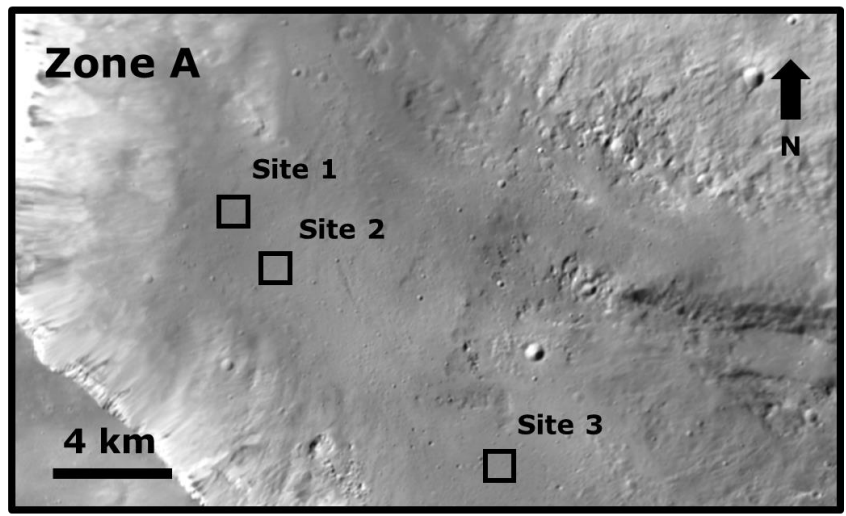

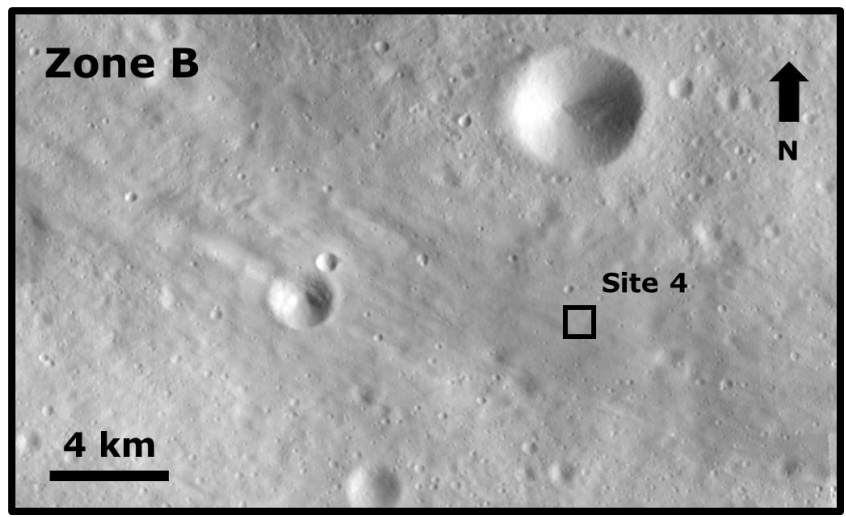

**Figure 4.** Surface dielectric model of each calibration site. Surface temperatures (left column) were retrieved from four VIR thermal images that overlapped the suggested calibration sites. Each thermal image was acquired in the Vestan morning during High-Altitude Mapping Orbit at a resolution of ~170 m/pixel. In order from top to bottom, the central coordinate of each VIR image is: (3°N, 151°E), containing Sites 1-3; (6°N, 158°E), containing Site 1; (23°S, 268°E), containing Site 4; and (18°S, 266°E), containing Site 4. Corresponding estimates of $\varepsilon'$ and tan $\delta$ (right column) are calculated for each VIR temperature image with a surface density of 1.30 g cm$^{-3}$. If local surface density is as high as 1.32 g cm$^{-3}$ in Sites 1-3, as indicated by regionally higher thermal inertia, our dielectric model suggests that $\varepsilon'$ is only ~1% greater than the surface average at the same temperatures.

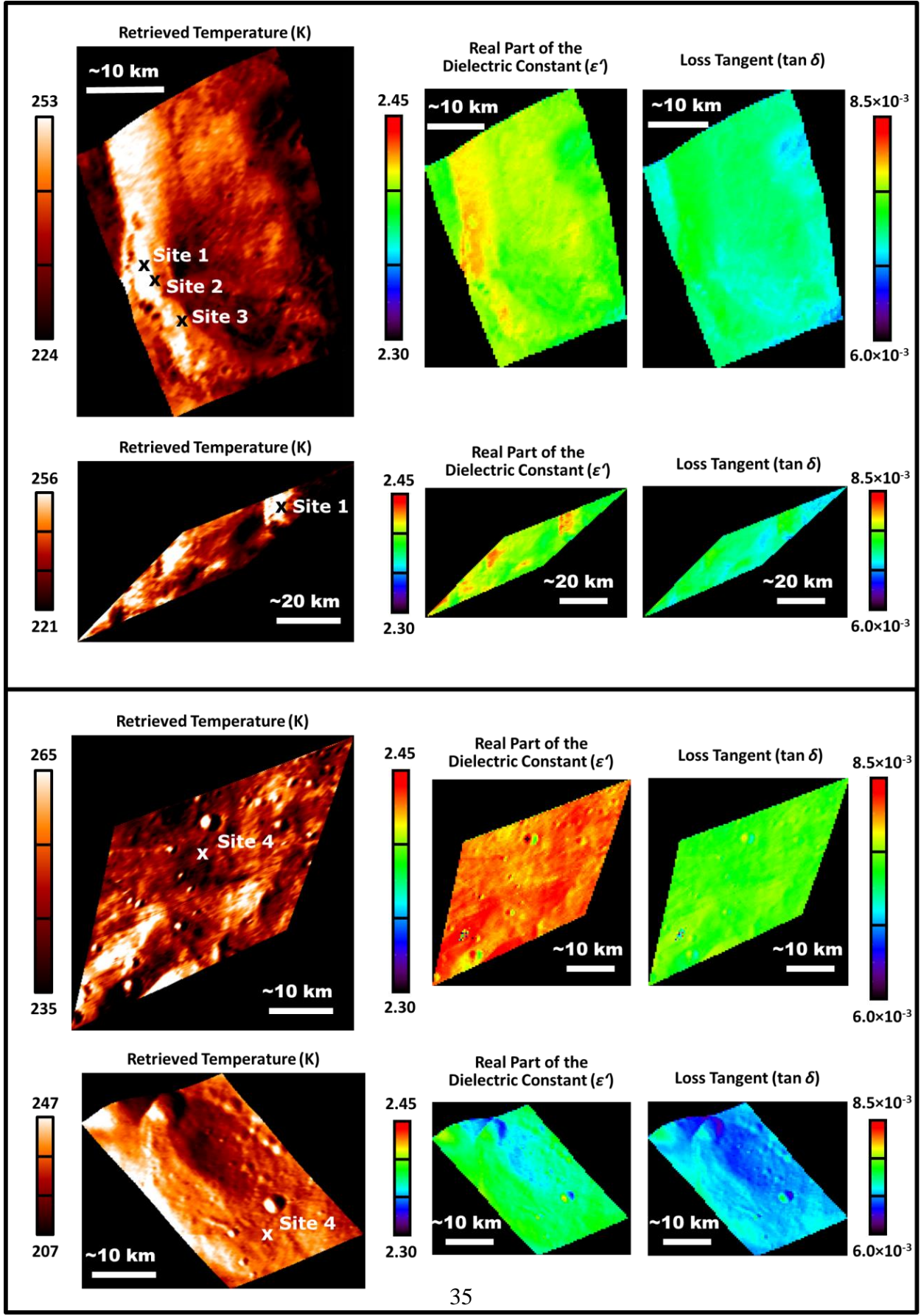